\documentclass[aps,twocolumn,showpacs]{revtex4}
\usepackage{graphicx}
\usepackage{bm}
\usepackage{amsmath}
\usepackage{amssymb}

\begin{document}

\title{Coherence scale of coupled Anderson impurities}
\author{Lijun Zhu}
\affiliation{Theoretical Division and Center for Nonlinear Studies,
Los Alamos National Laboratory,
Los Alamos, New Mexico 87545, USA}
\author{Jian-Xin Zhu}
\affiliation{Theoretical Division
and Center for Nonlinear Studies,
Los Alamos National Laboratory,
Los Alamos, New Mexico 87545, USA}
\date{\today}
\begin{abstract}

For two coupled Anderson impurities, two energy scales are present to characterize the evolution from local moment state of the impurities to either of the inter-impurity singlet or the Kondo singlet ground states. The high energy scale is found to deviate from the single-ion Kondo temperature and rather scales as Ruderman-Kittel-Kasuya-Yosida (RKKY) interaction when it becomes dominant. We find that the scaling behavior and the associated physical properties of this scale are consistent with those of a coherence scale defined in heavy fermion systems. 

\end{abstract}
\maketitle

\section{Introduction}
\label{sec:intro}

There has been much recent debate on the nature of a coherence scale for heavy fermion systems, which characterizes the evolution of $f$-electrons from the localized magnetic moments to itinerant quasiparticles~\cite{Pines08, footnote1}. In experiments this scale is manifested as the temperature $T^*$ above which the quasiparticle signatures vanish, for example, the Drude peak in the optical conductivity disappears, the magnetic entropy becomes $R\ln2$, and the resistivity (from magnetic ions) reaches a maximum value.  Traditionally, this is associated with the single-ion Kondo temperature $T_K$ which characterizes the Kondo renormalization of $f$-electrons to quasiparticles by forming spin singlet resonance state with conduction electrons. It is not known {\it a priori} that such a single impurity picture is applicable to the lattice, where magnetic ions are coupled by intersite interactions. This can be examined, for instance, by tuning the concentrations of magnetic ions in a parent non-magnetic compound while the dilute and dense limits provide information on the single-ion Kondo behavior and the Kondo lattice behavior with coupled local moments, respectively. One of such studies is on Ce$_x$La$_{1-x}$CoIn$_5$ and it is found that $T^*$ increases from the Ce-dilute limit with the increase of the concentration $x$ and can become one order of magnitude larger in the dense limit~\cite{Nakatsuji02}. An extensive analysis on a group of heavy fermion compounds displaying quantum critical properties~\cite{Pines08} shows that $T^*$ for these materials are indeed bigger than $T_K$. It is further discovered that $T^*$ scales as $\rho_0 J_K^2$, where $\rho_0$ is the density of states of conduction electrons at the Fermi energy and $J_K$ is the onsite Kondo coupling between spins of magnetic ions and conduction electrons. This is the scaling form for the intersite spin exchange interaction, namely, Ruderman-Kittel-Kasuya-Yosida (RKKY) interaction $I$, rather than the single-ion Kondo temperature $T_K \sim (1/\rho_0) e^{-1/(\rho_0J_K)}$.

To resolve the discrepancy between $T^*$ and $T_K$ therefore relies on the understanding of the interplay between the intersite coupling and the onsite Kondo coupling. While the associated lattice models for heavy fermion systems are in general difficult to solve, the two-impurity Anderson model (or the equivalent two-impurity Kondo model) presents such a competition effect in an exactly solvable way. With the assistance of the cluster dynamical mean-field theory (DMFT), a self-consistently solved two-impurity model also provides a solution to the Anderson lattice model~\cite{Kotliar05}. In the two-impurity Anderson (or Kondo) model, an antiferromagnetic RKKY interaction between two local moments favors an inter-impurity spin singlet state which competes with the Kondo singlet state, which is similar to the same type of competition in heavy fermion systems~\cite{Jayaprakash81,Jones87}.  It is known that for ferromagnetic RKKY interactions and a range of antiferromagnetic RKKY interaction, the evolution from the (high energy) local moment state to the (low energy) Kondo resonance state or the inter-impurity singlet state is through a two-stage process, characterized by two energy scales. Previous studies on this model were focused on the low energy scale, which characterizes the quantum phase transition or crossover between the two types of ground states~\cite{Jones87,Sakai89,Jones89,Affleck92,Sire93,Fye94,Gan95,Silva96,Zhu06}. The high energy scale, however, receives little attention hitherto, which is the focus of this study. While there are indications that it may be related to the single-ion Kondo temperature or RKKY interaction scale, there are no systematic studies on its scaling behavior and its associated physical properties, especially, when RKKY interaction is much bigger than the single-ion Kondo temperature. The reason to carry out this study is that, we show, this high energy scale shares the same properties as the coherence scale defined for the heavy fermion systems. Following the experimental definition, we denote this high energy scale as the coherence scale for coupled Anderson(Kondo) impurities~\cite{footnote1}. 

Besides the difference in the focus on the coherence scale, our study differs from previous ones in the following aspects. 1) We calculate the $T=0$ uniform and staggered spin susceptibilities in addition to the spectral functions (the imaginary parts of the single-particle Green's functions). We carry out an extended analysis on these dynamical quantities, for instance, a scaling analysis, and employ different quantities to identify and determine the characteristic energy scales. Therefore, we provide a detailed account for the physical properties associated with these energy scales. 2) We calculate these dynamical quantities with the recently-developed complete-Fock-space numerical renormalization group method~\cite{Anders06}. This method was developed to overcome the spectral weight loss in the intermediate and high energy range from traditional patching scheme in dynamical quantity calculations~\cite{Bulla08}, which becomes severe in the two-impurity problem with a larger eigen space. This approach is essential for determining the coherence scale which falls in this energy range. In particular,  we show by a comparison that the results from the patching scheme may lead to a misleading understanding of the coherence scale.  To our best knowledge, this is also the first practice to generalize this method to multi-impurity models. 3) It is also important to choose a system for which the single-ion Kondo temperature $T_K$ and the inter-impurity spin exchange interaction $I$ can be easily determined or directly given, which the identified characteristic energy scales can be compared to. We present results for two systems where two impurities are located far away from each other and at nearest neighbors on a three dimensional cubic lattice.  For two impurities located far away, RKKY interaction, which is generated from the virtual exchange of conduction electrons, is found to vanish. Theoretically, we can add a direct spin exchange term to simulate RKKY interaction effect~\cite{Jones87,Sakai89}. This system has been studied earlier, and we repeat the calculations on this system because it provides a reference system for the single-ion Kondo physics where $T_K$ can be determined.  We also provide new results for $I \gg T_K$. For two impurities sitting on nearest neighbors, a finite antiferromagnetic RKKY interaction between two impurities is found to be generated perturbatively in the order of $\rho_0 J_K^2$,  and is antiferromagnetic. As $T_K$ and $I$ are tuned in the same fashion as in realistic heavy fermion systems, this system provides a direct comparison to the experiments on the scaling behaviors of the characteristic energy scales. It also reveals an origin for a parity-splitting term which is inherent to the heavy fermion systems. 

From our numerical studies on the above two systems, we can in general identify two characteristic energy scales for two coupled Anderson impurities. While the ground state is found to be always a Fermi liquid fixed point, either a Kondo resonance state or the inter-impurity singlet state controlled by RKKY interaction, the low energy scale $T_L$ serves as the (local) Fermi liquid scale where Fermi liquid behaviors emerge. The high energy scale $T_H$ is found to be a spin fluctuation scale, where the imaginary parts of the spin susceptibilities reach their maximum values: the origin can be the Kondo spin-flip scattering or inter-impurity singlet to triplet excitation. In either case, the single-particle (charge) excitation gains considerable weight. We further find that $T_H$ for both systems has the same behavior: it increases from $T_K$ and then becomes $I$ when $I/T_K$ increases. This can be explained by whether the Kondo renormalization can reach $T_K$ or is already cutoff by the inter-impurity singlet to triplet excitation gap. Our results indicate that the physical properties above $T_H$ are still determined by the single-ion Kondo physics, which allows us to make an argument that $T_H$ can be generalized to the lattice system.  $T_H \approx I$ and its physical properties imply that it is consistent with the coherence scale $T^*$ defined for heavy fermion systems.  For completeness, we also show the behavior of $T_L$, which is the characteristic energy scale for the phase transition or crossover and is consistent with previous studies.  $T_L$ has different behaviors in the two considered systems due to two different origins, the proximity to a degeneracy point of the two competing grounds states and a parity-splitted quasiparticle hopping term. 

The rest of the paper is arranged as follows. In Sec.~\ref{sec:model}, we introduce the two-impurity Anderson model and the numerical method we adopted. The results for the direct and generated RKKY interaction cases are shown in Sec.~\ref{sec:twoscale}, where two energy scales are identified and their properties are characterized. We focus our discussions on the high energy scale, in Sec.~\ref{sec:lattice}, where we show that the scaling behavior of this scale and its associated physical properties are consistent with those of the coherence scale identified in the heavy fermion systems. In Appendix, we provide some technical details on how to evaluate $T_K$ and RKKY interaction. 

\section{Model and Method}
\label{sec:model}

The Hamiltonian for the two-impurity Anderson model can be written as
\begin{eqnarray}
H &=&   \sum_{{\bf k}\sigma} \epsilon_{\bf k} c^\dag_{{\bf k}\sigma} c_{{\bf k}\sigma} +{1\over\sqrt{N_c}}\sum_{{\bf k}\sigma i} \left({V_{\bf k} } e^{i {\bf k}\cdot {\bf r}_i}
c^\dag_{{\bf k}\sigma}f_{i\sigma} + \text{h.c.}\right) \nonumber \\
&+&  \sum_{i\sigma} \epsilon_f  f^\dag_{i\sigma} f_{i\sigma} + \sum_i U n_{f i\uparrow}n_{f i\downarrow} + I {\bf S}_1 \cdot {\bf S}_2 \;,
\label{eq:hamiltonian}
\end{eqnarray}
where $i$ sums over two impurity sites. This describes two interacting local orbitals  $f_{i\sigma}$ (Anderson impurities) embedded in a non-interacting conduction electron medium $c_{{\bf k}\sigma}$ with the system size $N_c$ and in hybridization with $c_{{\bf k}\sigma}$ with the strength $V_{\bf k}$ at each impurity site ${\bf r}_i$.  $\epsilon_f$ is the local orbital energy level and $U$ is the onsite Coulomb interaction for the impurities. $I$ is a direct spin exchange interaction between two impurity spins. In reality, such term, known as RKKY interaction, is perturbatively generated through a virtual process exchanging conduction electrons.  In the single occupancy limit of each orbital, this model can be mapped into a two-impurity Kondo model (some details are presented in Appendix).  

By taking the even or odd parity combinations of the local orbitals $f_{p=(e,o),\sigma} = (f_{1\sigma}\pm f_{2\sigma})/\sqrt{2}$, the fluctuations due to conduction electrons can be represented by two separate baths with the hybridization functions
\begin{eqnarray}
\Gamma_{e,o}(\omega)  &= &  -{1\over 2N_c} \text{Im} \left[ \sum_{\bf k} {V^2_{\bf k} |e^{i{\bf k}\cdot {\bf r}_1} \pm e^{i{\bf k}\cdot {\bf r}_2}|^2 \over \omega -\epsilon_{\bf k} + i 0^+} \right]\;.
\label{eq:hybfunc}
\end{eqnarray}
In our calculation we assume $V_{\bf k}=V$ and a three dimensional tight-binding dispersion $\epsilon_{\bf k} = - (D/3) \sum_{i=1}^3 \cos k_i a$ for conduction electrons with $D$ the half-bandwidth and $a$ the lattice constant.  We consider two cases with impurities located far away [case (i)] and at nearest neighbor lattice sites [case (ii)], where $|{\bf r}_1-{\bf r}_2|=\infty$ and $a$, respectively. $\Gamma_{e,o}(\omega)$ for these two cases can therefore be determined with Eq.~(\ref{eq:hybfunc}). Subsequently we can determine the Kondo coupling $J_K$ and the generated RKKY interaction $I$ while the calculation details are presented in Appendix~\ref{sec:appendix}.  We find that the hybridization functions for case (i) and case (ii) can be written as $\Gamma_{e,o}(\omega) = \Gamma_0$ and $\Gamma_{e,o}(\omega) = \Gamma_0(1\mp \omega)$, respectively, where $\Gamma_0 = \pi \rho_0 V^2$.  We take $U=2D$ and $\epsilon_f =-U/2$ in both cases. The Kondo coupling constant is found to be $\rho_0J_K = 8 \Gamma_0 / (\pi U)$. The generated RKKY interaction vanishes in case (i), indicating that it is effectively a single-ion Kondo problem, from which we can determine $T_K$. Theoretically, we add a direct spin exchange term $I {\bf S}_1 \cdot {\bf S}_2$ to simulate RKKY interaction. In case (ii), an antiferromagnetic RKKY interaction is perturbatively generated, which is determined to be $I\approx0.20\rho_0 J_K^2$ (see Appendix~\ref{sec:appendix}). Therefore we no longer need to add a direct spin exchange term. The ratio $I/T_K$ can be tuned by varying the Kondo coupling since $I$ and $T_K$ have different dependencies on $J_K$.  In our calculation, this is achieved by varying $V$ as $J_K\sim V^2/U$.

This impurity problem is amenable to the numerical renormalization group (NRG) method~\cite{Bulla08}, which prescribes a non-perturbative procedure to capture the low energy properties. It discretizes the hybridization function into a chain of electron orbitals with the energy decreasing in logarithmic scale.  In the two-impurity problem, as protected by the parity symmetry, $\Gamma_e$ and $\Gamma_o$ can be discretized separately. While only the head site of the chain couples to the impurities, we can solve the original model by an iterative diagonalization to incorporate gradually more low energy sites. In practice, as the eigen space increases with more sites included in each iteration, only certain number of low energy states are kept for the next iteration. From these NRG eigenstates, we can calculate dynamical quantities, $\Pi_{AB} (\omega) = -i \int_0^\infty d t e^{i\omega t}  \langle [ A(t), B(0)]_\pm \rangle$, in particular, the Green's function $G_{fp\sigma}$ with $A=B^\dag=f_{p\sigma}$, the uniform  ($\chi_u$) and staggered  ($\chi_a$) impurity spin susceptibilities, with $A=B = (S_{1z}+S_{2z})/\sqrt{2}$ and $(S_{1z}-S_{2z})/\sqrt{2}$, respectively. Traditionally, these quantities are calculated from the kept states from all iterations. However, kept states from different iterations are not necessarily orthogonal to each other. A patching scheme is adopted to avoid the overlap contributions, for instance, in Ref.~\cite{Sakai89}. This is to choose a weight function for the overlap spectrum between different iterations. The arbitrariness of the weight function leads to the inaccuracy of these dynamical quantities. The recently developed complete-Fock-space (CFS) method~\cite{Anders06} overcomes this problem, which we adopt in our calculation. It instead calculates dynamical quantities from all the discarded states (including all states from the final iteration), which is found to form an orthogonal Fock space conserving the total spectral density. Although the calculation time almost doubles with an additional backward iteration to determine the reduced density matrix, we find that the improvement in dynamical quantities is significant, especially in the energy range near and above $T_K$, which is of interest to our study. We will show a comparison for these methods in Fig.~\ref{fig:specomp}. While we typically choose a discretization parameter $\Lambda=2$ and keep 4000 lowest energy states in each iteration, we find that by choosing a smaller $\Lambda$ or keeping more states, the dynamical quantities have little change in the high energy range. 

\section{Emergence of two energy scales}
\label{sec:twoscale}

\subsection{Case (i): two impurities located far away from each other with a direct spin exchange interaction}

We first consider case (i), two impurities located far away from each other which can be represented by $\Gamma_e(\omega) = \Gamma_o(\omega)=\Gamma_0$. Although this case has been studied earlier~\cite{Sakai89}, we re-examine it here for two considerations. Firstly, as shown below, it provides a reference system to the single-ion Kondo physics when the direct spin exchange term is not added. $T_K$ can be determined. Secondly, it is a rare system exhibiting a quantum critical point with the added direct spin exchange term, with or without particle-hole symmetry (this is related to the vanishing parity-splitting potential scattering term). 

\begin{figure}[tbh]
\includegraphics[width=\columnwidth]{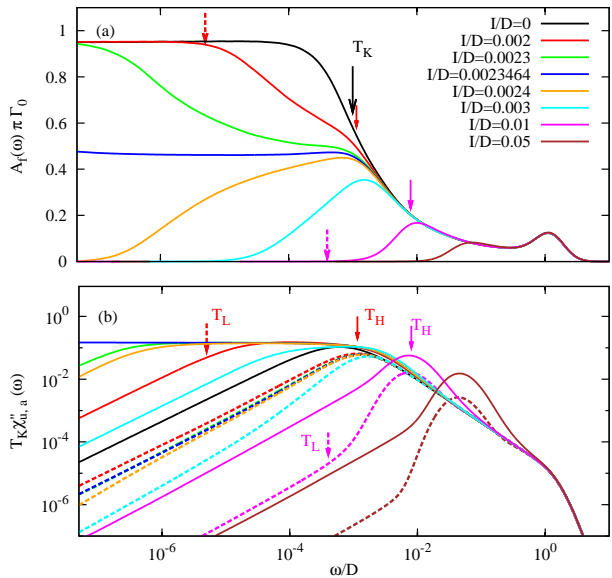}
\caption{(color online) Spectral functions (a), imaginary parts of the spin susceptibilities (b) as functions of energy for various values of  $I$  in case (i), two impurities located far away but coupled with a direct spin exchange interaction $I$. $\chi_u''$ and $\chi''_a$ in (b) are represented by dotted and solid lines, respectively.  Here $\Gamma_e = \Gamma_o=0.045\pi D$ and $\epsilon_f = -U/2=-D$. While only the $\omega>0$ range is shown, it is found that $A_{f}(-\omega) = A_{f}(\omega)$ and $\chi''(-\omega)=-\chi''(\omega)$.  $A_f$ is also symmetric for different parities and spins in this case. Two energy scales $T_H$ and $T_L$ (see definitions in the text) are illustrated for $I/D=0.002$ and $0.01$.}
\label{fig:phsym}
\end{figure}

The results of the spectral function $A_f(\omega) = - \text{Im} G_f(\omega)/\pi$, imaginary parts of the uniform ($\chi_u''$) and staggered ($\chi_a ''$) spin susceptibilities are shown in Fig. \ref{fig:phsym}. We choose a particle-hole symmetric case $\epsilon_f = -U/2 = -D$, and $\Gamma_{e,o}=0.045\pi D$ for $|\omega| \leq D$. Here $D=1$ serves as the energy unit. The unit for spin susceptibilities $(g\mu_B)^2$ is also set to 1, where $g$ and $\mu_B$ are the Land\'{e} factor and the Bohr magneton, respectively.  

In this case, without the direct spin exchange interaction $I$, we find that $\chi_u(\omega) = \chi_a(\omega)$, i.e., there is no correlation between two impurities $\langle S_{1z}S_{2z}\rangle=0$. Therefore, the spin dynamics, which controls the low energy properties, is the same as in the single-impurity Anderson model. In finite $U$ case, the Anderson model differs from its Kondo counterpart in additional features in high energies, i.e., the incoherent peaks of $A_f(\omega)$ at $\omega \approx \epsilon_f$ and $\epsilon_f +U$ corresponding to the free-orbital fixed point. In general, there exists an additional energy scale characterizing the evolution from the free-orbital fixed point to (or close to) the local moment fixed-point, where charge fluctuations freeze and $A_f(\omega)$ is small. In the Kondo limit, $U\to\infty$, these two fixed points can be thought of being pushed to the infinite energy. The low energy physics is controlled by the Kondo renormalization from the local moment fixed-point to the strongly-coupling Kondo fixed point, and can be characterized by a single energy scale, the single-ion Kondo temperature $T_K$. Below $T_K$,  $A_f$ has a resonance peak centered at the Fermi energy. [$ A_f(0)=1/(\pi \Gamma_0)$ by Friedel's sum rule. Our result has a few percent deviation which is due to the truncation of eigen space in NRG. In practice, we find that the improvement for this low energy sum rule can be achieved by keeping more states in NRG iterations and/or adopting a two-particle Green's function method~\cite{Bulla97}]. $T_K$ is also a spin-fluctuation scale, indicating the spin fluctuations, originated from the Kondo spin-flip scatterings, reach the maximal strength. As the $\omega=0$ Kondo fixed point is a Fermi lqiuid fixed point with scattering phase shift $\delta_{e,o} = \pi/2$, (local) Fermi liquid behaviors, such as $A_f(\omega) - A_f(0) \sim - \omega^2 $ and $\chi''_{u,a}(\omega) \sim \omega$, emerge when $|\omega| < T_F$ ($T_F$ is smaller than $T_K$ but is not an independent scale as it is proportional to $T_K$). We determine $T_K$ from the real part of the spin susceptibilities, $\chi_{u,a}'(\omega=0) \equiv 1/(4T_K)$, which can be determined from the imaginary parts with the Kramers-Kronig relation or the Korringa relation~\cite{Shiba75}: $\lim_{\omega\to 0} [\chi''_{u,a}(\omega)/\omega] = C_{u,a}= 2\pi [\chi_{u,a}'(0)]^2$. We find that $\chi_{u,a}'(0) = 2.5\times 10^2$, and $2.7\times 10^2$ by these two relations, and we determine  $\rho_0 T_K \approx  1.0\times 10^{-3}$. 

Upon adding an antiferromagnetic spin exchange interaction $I$, the low energy properties have changed: the spectral function is reduced from an energy scale around $T_K$, then then either increases again to the full weight $1/(\pi\Gamma_0)$ or decreases to 0 [a pseudogap state with $A_f(\omega) \sim \omega^2$, verified by a log-log plot] at a lower energy scale. The abrupt spectral weight change at the Fermi energy is a signature of the two-impurity quantum phase transition~\cite{Jones87,Sakai89}:  the ground state changes from a Kondo resonance states with two impurity spins forming Kondo singlets with the conduction electrons, to an inter-impurity spin singlet states with two impurity spins forming singlets by themselves and being decoupled from the conduction electrons. The critical values is found to be  $I_c \approx 2.3 T_K$  which is also consistent with previous studies~\cite{Jones87,Sakai89}.  Apparently, the single-ion Kondo picture with a single scale, $T_K$, no longer applies and two energy scales are needed to explain the dynamical quantities. On both sides of the critical point, Fermi liquid behaviors persist at low energies, such as $\chi''_{u,a}(\omega) = C_{u,a} \omega$. We therefore associate the low energy scale $T_L$ as the Fermion liquid temperature.  But we need to pay attention that $T_L$ alone cannot differentiate the Kondo resonance state and the inter-impurity singlet state, both of which are Fermi liquid fixed points but have different spectral weights  or the scattering phase shift at the Fermi energy. The high energy scale $T_H$ is associated with a spin-fluctation scale, where $\chi_{u,a}''(\omega)$ reach the maximal strength (for $I$ close to $I_c$, where $\chi''_a(\omega)$ becomes relatively a constant, $T_H$ is the high energy edge). 

\begin{figure}[tbh]
\includegraphics[width=\columnwidth]{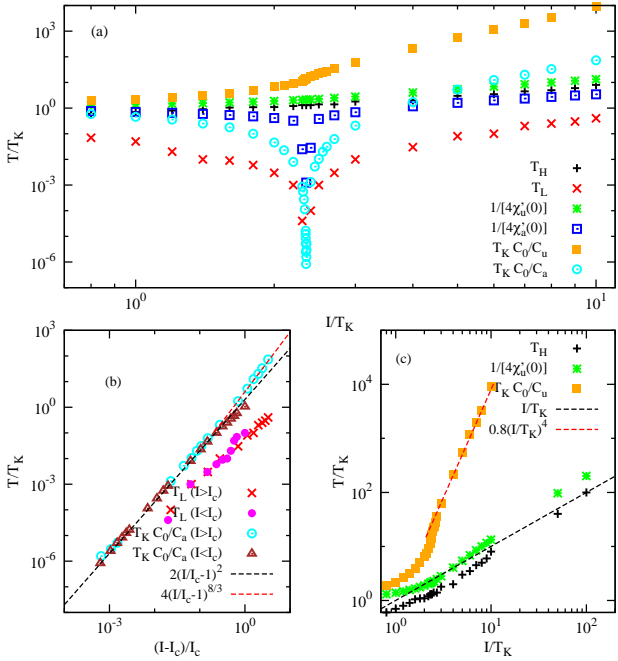}
\caption{(color online) (a) Characteristic energy scales as functions of $I/T_K$. The energy scales determined from $1/\chi_{u,a}'(0)$, and $C_{u,a}=\lim_{\omega\to0} \chi''_{u,a}(\omega)/\omega$ are also shown. (b) shows the low energy scale as a function of $(I-I_c)/I_c$. Two dotted lines are fitting lines $T/T_K = 2.0 (I/I_c-1)^2$ and $T/T_K = 4.0(I/I_c-1)^{8/3}$. (c) shows amplified region of the high energy scale. Two dotted lines are the guide line $T=I$ (black) and a fitting line $T/T_K =0.8 (I/T_K)^4$ for $1/C_u$.  }
\label{fig:energyscale}
\end{figure}

According to this observation, we define $T_H$ as the peak position of $\chi''_u(\omega)$ and $T_L$ as the onset energy when $\chi''_{u,a}(\omega)$ becomes linear in $\omega$. The former can be easier determined from the numerical data. The errors come from systematic ones such as numerical discretization and the broadening of the delta function to a log-Gaussian form. The latter, however,  has arbitrariness because the crossover to the Fermi liquid behaviors spans a wide energy range. Therefore, we also provide characterizations from $1/[4\chi'_{u,a}(0)]$ and $C_{u,a}$. In the single-ion case, $1/[4\chi'_{u,a}(0)] = T_K$ while $C_{u,a}=C_0 \sim 1/T_K^2$. When $I$ is close to $I_c$, we find that the constant piece in $\chi''_a(\omega)$ remains relatively unchanged, implying $C_a T_L = C_0 T_K$. In this region, $T_K C_0/C_a$ provides a more reliable method to determine $T_L$. We plot the energy scales determined from these methods in Fig.~\ref{fig:energyscale}. 

\begin{figure}[tbh]
\includegraphics[width=\columnwidth]{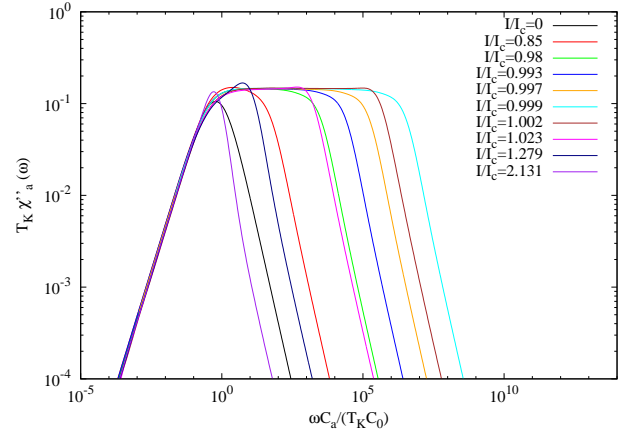}
\caption{(color online) Scaling analysis on the staggered spin susceptibility. $\chi''_a(\omega)$ as functions of $\omega$ are shown for different $I$s, as in Fig.~\ref{fig:phsym}(b). Here $\omega$ is scaled with $T_K C_0/C_a$ to the scaling behavior. }
\label{fig:scaling}
\end{figure}

From Fig.~\ref{fig:energyscale},  we learn that $T_L$ and $T_K C_0/C_a$ have similar behaviors (the magnitude difference is similar to the difference in $T_F$ and $T_K$ in the single-ion case). As $T_K C_0/C_u$ is determined directly from numerical data, it shows better asymptotic behaviors. Near $I_c$, we find that it can be fitted as $(I-I_c)^2$. This indicates a uniformly vanishing energy scale, which characterizes the continuous quantum phase transition. The exponent 2 is also consistent with previous studies~\cite{Jones87,Sakai89}. But away from $I_c$, the exponent is smaller or bigger than 2 for $I\to0 $ and $I \gg I_c$ respectively.  In the latter case, it is rather fitted as $(I-I_c)^{8/3}$ within a range of $I-I_c>0.2I_c$.  This fractional exponent may not be universal, as we can also fit $1/C_a \sim 1/C_u \sim I^4$ for $I \gg I_c$. Another check for the low energy scale is from a scaling analysis, by rescaling the energy with respect to $1/C_a$. We carry on such an analysis for $\chi''_a(\omega)$, which is  shown in Fig.~\ref{fig:scaling}. Once the energy is rescaled with $T_K C_0/C_a$,  the low energy part of  $\chi''_a(\omega)$ falls in the same universal curve within $|\delta I/I_c|  < 20\%$, which is associated with the universality of the two-impurity quantum critical point. Indeed, $T_K C_0/C_a$ provides a faithful representation of the low energy scale $T_L$. On the other hand, the Korringa relation is violated for finite $I$s. While $1/\chi'_a(0)$ has the same trend as $T_L$, it rather vanishes  logarithmically $\sim 1/\ln |I-I_c|$.  

$T_H$ and $1/[4\chi'_u(0)]$ have similar behaviors: they increase uniformly when $I$ increases. $T_K C_0/C_u$ follows the same trend, but increases more rapidly when $I>I_c$.  Clearly, $T_H$ increases from $T_K$ and becomes the RKKY scale $I$ when $I \gg T_K$.

\subsection{Case (ii): two impurities sitting on nearest neighboring sites with generated RKKY interaction}

\begin{figure}[tbh]
\begin{center}
\includegraphics[width=0.8\columnwidth]{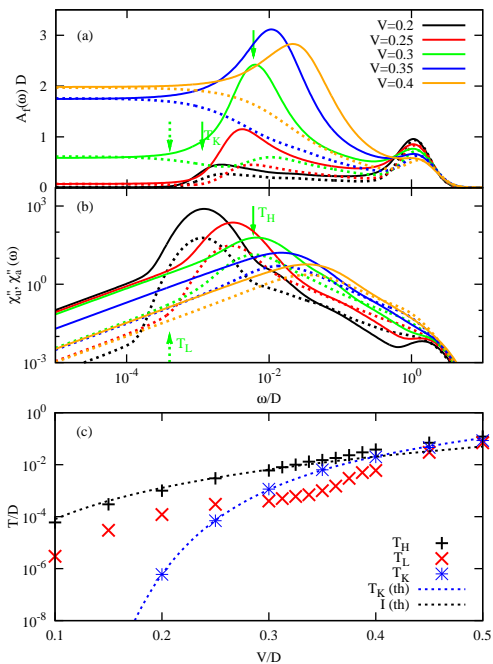}
\end{center}
\caption{(color online) Spectral functions (a), imaginary parts of the spin susceptibilities (b) as functions of energy for various values of hybridization constant $V$ in case (ii), two impurities sitting on nearest neighbors with generated RKKY interaction. The solid (dashed) lines represent the the even (odd) channels for $A_f$, and $\chi_a''$ ($\chi_u''$) for spin susceptibilities.  With $\Gamma_e(-\omega) = \Gamma_o(\omega)$, $A_{e,o}(\omega) = A_{o,e}(-\omega)$ is satisfied. (c) shows different energy scales as functions of $V$ (in unit of $D$).  $T_K$ is determined from calculations by assuming $\Gamma_{e,o}=\Gamma_0$,  and is fitted (blue dashed line) by $\rho_0 T_K= 0.5  \exp[- (1/\rho_0 J_K) + (1/2) \ln (\rho_0 J_K) + 1.58(\rho_0 J_K)^2]$. The black dashed line is the RKKY scale $I=0.20\rho_0 J_K^2$.}
\label{fig:phasym}
\end{figure}

We proceed to study case (ii), two impurities sitting on nearest neighbors which can be represented by $\Gamma_{e,o}(\omega) = \Gamma_0 (1 \mp \omega)$. The results are shown in Fig.~\ref{fig:phasym}. We observe similar behaviors, where small (large) $V$ cases can be compared with large (small) $I/T_K$ cases in case (i). The differences are, $A_f$ for even and odd parties split due to different $\Gamma_{e,o}(\omega)$, and $A_f(0)$ smoothly varies rather than a jump. We can also identify two similar energy scales in $A_f$ and $\chi_{u,a}''$ and plot them as functions of $V$ in Fig.~\ref{fig:phasym}(c). As learned from case (i), both $T_H$ and $T_L$ deviate from $T_K$ when RKKY interaction is finite, it is not possible to determine $T_K$ from the dynamical quantities in this case. Instead, we calculate $T_K$ from the corresponding single impurity case with $\Gamma_{e,o} (\omega)= \pi \rho_0 V^2$ as in case (i) with the same $V$. We find that it can be fitted by the standard expression $\rho_0 T_K= 0.5  \exp[- (1/\rho_0 J_K) + (1/2) \ln (\rho_0 J_K) + 1.58(\rho_0 J_K)^2]$ where $\rho_0 J_K = 8\rho_0 V^2/U$.  RKKY interaction is determined to be antiferromagnetic $I\approx 0.20\rho_0 J_K^2$ (see Appendix~\ref{sec:appendix}). As $V$ increases, both $I$ and $T_K$ increase monotonically but their ratio decreases. This provides a direct simulation to realistic heavy fermion materials; both follow the Doniach's phase diagram [cf. Fig.~\ref{fig:phasym}(c)]~\cite{Doniach77}. 

\begin{figure}[tb]
\includegraphics[width=\columnwidth]{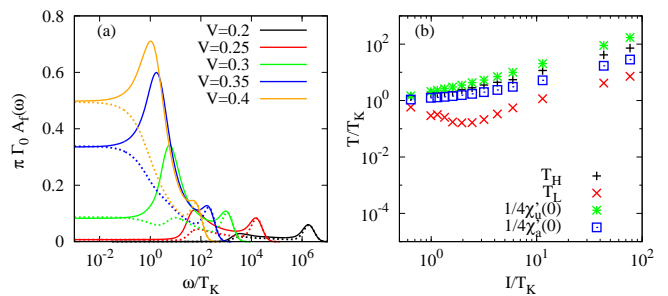}
\caption{(color online) The rescaled spectral functions (a) and energy scales (b) in case (ii), impurities sitting on nearest neighbors with generated RKKY interaction. These are the same plots as in Fig.~\ref{fig:phasym} (a) and (c). After a rescaling, they can be compared with Fig.~\ref{fig:phsym}(a) and Fig.~\ref{fig:energyscale}. }
\label{fig:tscale}
\end{figure}

Since $T_K$  and $\Gamma_0$ are varying as functions of $V$ in this case, it is helpful to replot our results in scaled forms. For instance, we can rescale energies with $T_K$ and $A_f$ with $1/(\pi \Gamma_0)$ for each value of $V$. This allows a direct comparison to case (i) where $T_K$ is fixed. The rescaled spectral functions and energy scales are plotted in Fig.~\ref{fig:tscale}.  For the spectral functions, it is interesting to observe that the high energy parts follow the universal Kondo curve, i.e., neglecting the incoherent peaks, this curve describes the Kondo renormalization from a local moment state at infinite energy to the low energy Kondo fixed point. However, this curve is rather associated with a particle-hole asymmetric case, or with a finite potential scattering term. The potential scattering terms for the even and odd parities have the same magnitude and are opposite in sign. The potential scattering term in the single-ion model is known not to change the universality of the Kondo fixed point: it simply shifts the resonance peak position and contributes an additional phase shift $\pi/2 + \delta_p$. Therefore, $A_f(0)$ is reduced from $1/(\pi\Gamma_0)$. The $V=0.4D$ case with $I \ll T_K$ presents such an example. When $V$ decreases, $I/T_K$ increases, and the low energy physics begins to be affected by RKKY interaction. Similar to the particle-hole asymmetric two-impurity model studied before: the system shows a crossover from the Kondo resonance state to the inter-impurity spin singlet state, rather than a continuous transition. $T_L$ remains finite near $I \approx 2.2 T_K$ (at $V/D \approx 0.35$). The divergence in $\chi_a'(0)$ is also absent.  Nevertheless, we find that $T_H$ has the same behavior as in case (i): it uniformly increases from $T_K$ to $I$ when $I/T_K$ increases.  

\subsection{Origin of two energy scales} 

The emergence of two energy scales indicates the effect of a finite RKKY interaction on the Kondo renormalization. While the triplet configuration of the two impurity spins, as an $S=1$ impurity, can undergo the spin-flip scattering to itself in favor of Kondo resonance, the singlet configuration only couples to the conduction electron by being firstly excited to the triplet configuration, through $({\bf S}_1-{\bf S}_2)\cdot (c_{e\sigma}^\dag {\vec \tau}_{\sigma\sigma'} c_{o\sigma} +c_{o\sigma}^\dag {\vec \tau}_{\sigma\sigma'} c_{e\sigma} )$, where $c_{p\sigma} \sim \sum_{\bf k} c_{{\bf k}\sigma} (e^{i {\bf k}\cdot {\bf r}_1} \pm  e^{i {\bf k}\cdot {\bf r}_2})$. RKKY interaction creates an excitation gap between the singlet and triplet configurations of the two impurity spins, causing these two configurations to renormalize differently. This can be evidenced from the difference in the uniform and staggered spin susceptibilities. When RKKY interaction is antiferromagnetic~\cite{footnote2} but small, though the singlet configuration is favored, the Kondo scattering between the singlet and triplet configurations, on the energy scale of $T_K$, is able to overcome the singlet-to-triplet gap and the Kondo resonance state persists at low energies.  When RKKY interaction is much larger than $T_K$,  the Kondo resonance is suppressed at low energies, leading to a state with gapped quasiparticle excitations, denoted as the inter-impurity singlet state.  

While both the Kondo resonance state and the inter-impurity singlet state belong to the Fermi liquid fixed points, $T_L$ characterizes the energy difference between these two states. Above $T_L$, the system evolves into an intermediate state where the inter-impurity singlet to triplet excitation and the Kondo scattering are on equal footing and the physical properties deviate from FL behaviors. This intermediate state is also the quantum critical state in case (i) when $T_L$ vanishes at $I=I_c$. From our results, we learn that the spectral weight for the quantum critical state is exactly the half of that for the Kondo resonance state and the staggered spin susceptibility diverges logarithmically. We find that these properties are consistent with the Majorana fermion picture proposed by the Bosonization approach~\cite{Sire93,Gan95}. This picture dictates that the quantum critical state is a partially screened Kondo state with only half of the impurity degrees of freedom as a Majorana fermion forming a resonance model with the extended electron degrees of freedom. This can also be evidenced from the thermodynamics calculated from NRG~\cite{Campo04} that the entropy (per impurity) for this state is $(\ln2)/2$. While $T_L \sim (I-I_c)^2/T_K$ in case (i) indicates the proximity to the degeneracy point between the Kondo resonance state and inter-impurity singlet state,  the finite $T_L$ in case (ii) is due to another effect. In case (ii), the particle-hole symmetry is actually broken in each channel due to the presence of a parity-splitting potential scattering term $V_{-} (c^\dag_{e\sigma}c_{e\sigma} - c^\dag_{o\sigma} c_{o\sigma})$. It is shown~\cite{Gan95,Affleck92} that such term will generate a coupling between the other half of the impurity degrees of freedom (the other flavor of Majorana fermion) and the extended degrees of freedom. Therefore, there is always a finite quasiparticle spectral weight at the Fermi energy.  The FL temperature is also finite $T_L\sim T_K(\rho_0V_-)^2$~\cite{Gan95,Affleck92,Affleck09}.  We notice that such a parity-splitting term can also be generated by other forms of particle-hole asymmetry, such as a hybridization (hopping) term $t_f(f^\dag_{1\sigma} f_{2\sigma} +h.c.)$~\cite{Sakai89}, the asymmetry in energy dependence of Kondo coupling~\cite{Silva96} or a regular potential scattering term~\cite{Zhu06}.  Compared with these studies,  case (ii) we studied here points out an origin for the parity-splitting term which is inherent to the heavy fermion systems, and cannot be tuned away. 

\section{The relation between $T_H$ and the coherence scale}
\label{sec:lattice}

While the properties of $T_L$ are the focus of previous studies on this model~\cite{Jones87,Sakai89,Jones89,Affleck92,Sire93,Fye94,Gan95,Silva96,Zhu06} and our results are in agreement with these studies, we focus our discussions on scaling behavior and the properties of $T_H$ which has received little attention.  

As shown in Figs.~\ref{fig:phsym}(c) and~\ref{fig:tscale}(b), $T_H$ has the same scaling behavior in the two cases studied above: $T_H \approx T_K$ (or $I$) for $I \ll T_K$ (or $\gg T_K$) and a value in-between when $I$ and $T_K$ are comparable. When $I \ll T_K$,  the Kondo effect dominates. A small RKKY interaction in this limit only enhances slightly the staggered spin fluctuations and affects the physical properties in the energy range $(T_K-I, T_K)$ while the Kondo renormalization above $T_K$ remains unaffected. Therefore,  $T_H\approx T_K$ in this limit. $T_K$ lies in the crossover region between the local moment and the strong coupling fixed points in the single-ion Kondo renormalization. Correspondingly, the imaginary parts of the spin susceptibilities change behaviors at $T_K$ from $\sim 1/\omega$ to $\sim \omega$, i.e,  the spin fluctuations reach the maximum strength associated with the spin-flip Kondo scattering.  The quasiparticle spectral weight also increases rapidly at $T_K$. Indeed, $T_H$ (being $T_K$) serves as the onset scale for the formation of coherent quasiparticles. We notice that quasiparticles at this energy scale do not necessarily exhibit Fermi liquid behaviors, which rather develop at a lower energy scale. When $I \gg T_K$, RKKY interaction suppresses the Kondo scattering from the Fermi energy up to the scale $I$. The Kondo renormalization is cutoff by $I$ before reaching $T_K$. The imaginary parts of the uniform and staggered spin susceptibilities reach the maximum values at $I$, but have a hump structure instead of the smooth crossover between the local moment to FL behaviors as in the Kondo resonance state. While this hump is due to the singlet to triplet spin fluctuations between two impurity spins, it is broadened by the Kondo scattering. In other words, impurities are still weakly coupled to the conduction electrons to form quasiparticles, i.e., the coupling is $J_K$ rather than $J_{K,eff} \to \infty$ at the Kondo fixed point. The spectral weight is still finite and reaches the maximum value at $I$ though it is small. $T_H$ is determined by $I$ in this limit, which can be understood as a cutoff effect on the Kondo renormalization. When $I$ is comparable with $T_K$,  the Kondo scattering and inter-impurity singlet to triplet excitation have comparable strengths. $T_H$ as the location for the maximum spin fluctuations lies in between $T_K$ and $I$ and increases when $I/T_K$ increases. In all these cases, we find that $T_H$ serves as the scale where the physical properties begin to be strongly influenced by the formation of quasiparticles. Therefore, we can associate $T_H$ to the coherence scale for two coupled impurities.  

We notice that the properties of $T_H$ are in agreement with those of the coherence $T^*$ in heavy fermion systems. For instance, the changes in optical conductivities, the tunneling spectroscopy, and the Hall effect are related to the change of the quasiparticle spectral weight while the magnetic responses and the scattering rate are affected by the strong spin fluctuations at this scale.  As observed from the two-impurity model, the single-ion Kondo temperature $T_K$ loses its manifestation in physical properties when $I>T_K$. This is also true for heavy fermion systems. In experiments, $T_K$ is determined by the dilute magnetic ions limit, as in Ref.~\cite{Nakatsuji02}. Indeed, it is in general one order of magnitude smaller than $T^*$~\cite{Pines08}. This implies that most heavy fermion compounds fall in the $I>T_K$ regime with magnetic orders developing at low temperatures. In this case, we find that $T_H$ is indeed determined by RKKY interaction, which is consistent with the experimental analysis~\cite{Pines08}. In this case, we notice that the quasiparticles at the scale $I$ are different from Landau quasiparticles displaying Fermi liquid behaviors: they rather are underscreened Kondo excitations with non-Fermi liquid behaviors. Nevertheless, the dynamical measurements determining the coherence scale indeed are not necessary in the Fermi liquid regime. Besides, Fermi liquid behaviors develop at a lower energy scale. We find that the Fermi liquid parameters, such as $C_{u,a} = \lim_{\omega\to0}[\chi''_{u,a}(\omega)/\omega]$, are also determined by $I$. In this sense, the coherence scale indeed appears in the Landau parameters, which are manifested also in static property measurements. This provides an explicit explanation to the scaling behavior of $T^*$ as RKKY interaction rather than the single-ion Kondo temperature, and it can be interpreted as a coherence scale.

\begin{figure}[tb]
\includegraphics[width=\columnwidth]{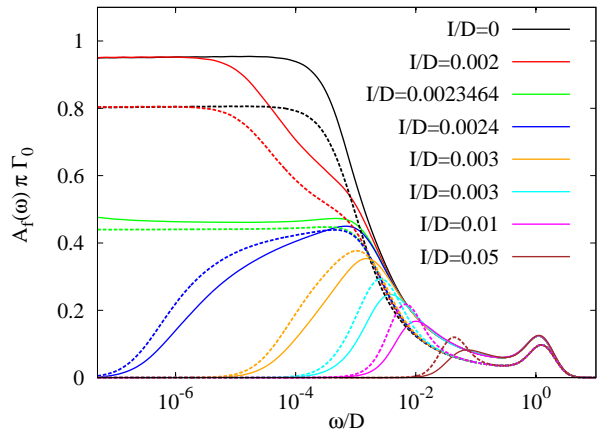}
\caption{(color online) Comparisons between the spectral functions obtained by CFS-NRG method (solid lines) and the patching scheme (dotted lines) (please refer to Sec.~\ref{sec:model} for details of these two methods). The parameters taken are the same as in case (i), or Fig.~\ref{fig:phsym}.  For both methods, the NRG iteration part is the same. A weight function $w(x)=x$ is commonly chosen for the patching scheme. The spectral functions obtained by CFS-NRG method have high weights in the whole energy range. We explicitly verify the sum rule $\int d\omega A_f(\omega)=1$, and find the deviation in practice is less than $0.1\%$. }
\label{fig:specomp}
\end{figure}

Then the question is whether $T_H$, the coherence scale for two Anderson impurities, can be generalized to the heavy fermion lattice systems. This relies on an explicit calculation on the lattice models, for instance, solving the periodic Anderson lattice model through a self-consistent two-impurity Anderson model with the cluster DMFT~\cite{Kotliar05}. In this approach, one chooses the neighboring two sites of the lattice as a cluster and treat the lattice as the repetitions of this cluster. The even and odd parity states therefore correspond to the degrees of freedom at momentum (0,0,0) and ($\pi,\pi,\pi$), respectively, for a three dimensional lattice. Once self-consistently solved, the two impurities have the same properties as the two sites in the lattice incorporating two-site correlations. Although the self-consistency procedure hasn't been done explicitly, we argue that the generalization of the coherence scale can be valid. The most important result from our calculations is that the high energy properties above $T_H$ are governed by the single impurity physics. i.e., the intersite coupling only modifies the low energy properties. This can be observed from the spectral functions [Cf. Fig. (1)a and Fig. 3(a)], which follow the universal single-ion Kondo curve at high energies but their low energy weights are suppressed systematically from the scale $T_H$, i.e., $T_H$ being RKKY interaction, is a cutoff scale on the single-ion Kondo renormalization. This statement, however, cannot be made from the patching scheme calculations. In Fig.~\ref{fig:specomp}, we show the spectral functions obtained from both CFS-NRG method and the patching scheme. While these methods may give the same positions for the characteristic energy scales, as the NRG iteration part is the same, the spectral function obtained by the patching scheme may lead to a different understanding on $T_H$ for $I \gg T_K$. We observe enhanced quasiparticle excitations at $\omega \approx I$ compared with the single-ion case. This is due to the choice of the weight function between the overlapping energy range, which, commonly, favors the states in later iterations (lower energies) and exaggerates its weight. However, this may indicate that, due to RKKY interaction $I$, considerable spectral weight transfers from the incoherent part to the scale $I$ to form coherent excitations, rather the cutoff effect indicated by the CFS-NRG result.  It is from the spectral functions that we determine the new Weiss fields to recalculate the two-impurity problem. As the high energy part above $T_H$ is determined by the single-ion Kondo physics, it remains unchanged in the self-consistency procedure. The low energy part, however, is expected to be modified in the self-consistency procedure. But as $I$ is the cutoff scale, it does not involve any spectral weight transfer from the incoherent part, or more strictly, from part with energy higher than $I$. Therefore, $T_H$ remains as the high energy scale for the lattice. A caveat for this argument is not to violate the spectral sum rule. As seen from our calculation, the spectral weight of the low energy part only consists a negligible portion of the total spectral weight, for instance, valid for Ce compounds. We notice that this argument may not hold for cases where the weight transfer between the incoherent part and coherent part is significant.

\section{Conclusion}
\label{sec:conclusion}

In summary, we have studied the two-impurity Anderson model with the complete-Fock space numerical renormalization group method. We find that due to the competition between the Kondo effect and RKKY interaction which is the spin exchange interaction between two impurities, there are in general two energy scales present. The high energy scale characterizes the quasiparticle formations and the low energy scale specifies when these quasiparticles develop Fermi liquid behaviors. In between, there exists an intermediate state with non-Fermi liquid behaviors where both the Kondo scattering and the excitations between the singlet and triplet configurations of the two impurity spins play equally important roles.  

The focus of our study is on the high energy scale which we associate with the coherence scale in heavy fermion systems. In the two-impurity model,  it marks the scale with maximum spin fluctuations. While the spin fluctuations can be due to the Kondo scattering with conduction electrons or the singlet to triplet excitations between the impurity spins, their relative strength is tuned by the ratio between RKKY interaction and the Kondo temperature. When the Kondo effect wins, the high energy scale is determined by the Kondo temperature, where quasiparticles begin to gain considerable weight. When RKKY interaction dominates, the high energy scale is determined by RKKY interaction, which can be understood as a cutoff on the Kondo renormalization by a spin excitation gap. The quasiparticle spectral weight therefore reaches the maximum value at this scale. In both cases, the high energy scale characterizes when the quasiparticles gain considerable weight to influence the physical properties. This is consistent with the definition of the coherence scale in heavy fermion systems. The most important result is that the physical properties above the high energy scales is determined by the single-ion Kondo physics and is expected to remain unchanged in a cluster DMFT procedure. This enables us to associate the high energy scale directly to the coherence scale of heavy fermion systems. Besides sharing the same physical properties, they also have a same scaling behavior. When RKKY interaction is much bigger than the single-ion Kondo temperature, the coherence scale is determined by RKKY interaction which is consistent with the experimental analysis~\cite{Pines08}. This provides an explicit theoretical description for such an energy scale.

The low energy scale, as the Fermi liquid temperature for quasiparticle, is expected to change in the cluster DMFT self-consistent procedure. This prevents a direct connection between the two-impurity quantum phase transition to the magnetic quantum phase transition in heavy fermion systems. However, we expect both origins for the low energy scale in the two-impurity model will play important roles in determining the low energy scale(s) in a lattice model. The degeneracy point between magnetic excitations and Kondo scattering also exists in a lattice, for which Doniach's criterion remains the same, i.e., $I_c/T_K$ is of order unity. If the two-impurity critical dynamics drives a magnetic transition for the lattice, it will be interesting to check the existence of Majorana fermion: many of its properties are in accordance with the local quantum critical behavior~\cite{Si08}. On the other hand, the two-impurity quantum critical point is unstable against certain forms of particle-hole asymmetry. For instance, the parity-splitting potential scattering term is always finite in case (ii) of our study, which is inherent to the heavy fermion lattices. In a lattice, this can be related to asymmetry between degrees of freedom at momentum (0,0,0) and ($\pi,\pi,\pi$). We have shown that it not only generates a finite Fermi liquid temperature but also generates a finite spectral weight at the chemical potential [cf. Fig.\ref{fig:phasym}(a)]. The parity-splitting term also relates to a finite $\text{Re}\Sigma_{12}(\omega=0)$, which corresponds to a kinetic energy term for $f$-electrons~\cite{Zhu06}. This might help to stabilize a finite effective Kondo scale. Incidentally, as the potential scattering term scales as the Kondo coupling, the associated (low) energy scale then also has the same scaling relation as RKKY interaction. 

We expect that the two-impurity Kondo physics will also help to understand the Mott transition and unconventional superconductivity in other strongly correlated systems. The Kondo physics has provided an elegant explanation to the Mott transition in the single-site DMFT approach to the Hubbard model~\cite{Georges96}. However, the intersite spin exchange interaction, similar to RKKY interaction, is absent in the single-site approach. We learn from the two-impurity model that such an intersite interaction competes with the local Kondo dynamics to modify the low energy properties. Indeed a multi-site cluster DMFT calculation shows different properties the metallic and insulating phases near the transition point~\cite{Park08}.

\acknowledgements
We would like to thank Qimiao Si, Joe D. Thompson and Chandra M. Varma for helpful discussions. One of us (LZ) is also grateful to Chandra M. Varma for collaboration on related research. This work was supported by the National Nuclear Security Administration of the U.S. DOE  at  LANL  under Contract No. DE-AC52-06NA25396, the U.S. DOE Office of Science, the LDRD Program at LANL, in part by the NSF under Grant No. PHY05-51164.  We also acknowledge the hospitality of Kavli Institute for Theoretical Physics (UCSB).

\appendix
\section{Determination of the hybridization functions and the interaction parameters}
\label{sec:appendix}

We explain some of the calculation details for the hybridization functions and the interaction parameters such as the Kondo coupling $J_K$ and RKKY interaction $I$.

\begin{figure}[tbh]
\includegraphics[width=\columnwidth]{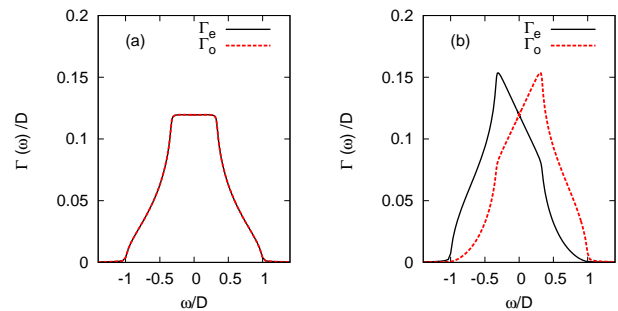}
\caption{(color online) The hybridization functions for the even and odd parity channels for two Anderson impurities in a three dimensional cubic lattice. (a) is for two impurities located far away from each other (here ${\bf r}_1 - {\bf r}_2 =40 a {\hat x} $) and (b) is for two impurities sitting on nearest neighbor sites (${\bf r}_1 - {\bf r}_2 =a {\hat x} $) . Here $V=0.3D$.}
\label{fig:gamma}
\end{figure}

The hybridizations functions can be calculated from Eq.~(\ref{eq:hybfunc}) by the given hybridization constant $V_{\bf k}$, the conduction electron dispersion $\epsilon_{\bf k}$, and the impurity locations. We choose $V_{\bf k}=V$, $\epsilon_{\bf k} = - (D/3) \sum_{i=1}^3 \cos k_i a$. For two impurities located far away from each other and sitting on nearest neighbors, we determine $\Gamma_{e,o}(\omega)$ numerically and the results are shown in Fig.~\ref{fig:gamma}. In Fig.~\ref{fig:gamma}(a), we find that $\Gamma_e(\omega) = \Gamma_o(\omega) = \pi V^2 \rho(\omega)$, where $\rho(\omega)$ is the density of state function for the dispersion $\epsilon_{\bf k}$. Near the Fermi energy, $\Gamma_{e,o}(\omega)$ are constants.  In Fig.~\ref{fig:gamma}(b), we find that $\Gamma_{e,o} (\omega) \sim 1\mp \omega$ near the Fermi energy. For simplicity, we take their low energy forms to the whole band. The difference is above the Van-Hove singularity at $|\omega| \approx 0.3D$ which doesn't affect the low energy physics. 

The Kondo coupling constant $J_K$ can be obtained from a canonical transformation from the Anderson model to the Kondo model. In the two-impurity Kondo model, as the Kondo scattering can involve electrons from the same or different parity channels, there are different types of Kondo exchange interaction  
\begin{equation} 
J_{pp'} (\omega, \omega')/D =  {1\over \pi} \left[\Gamma_p(\omega) \Gamma_{p'}(\omega')\right]^{1/2} C(\omega, \omega')\;, 
\end{equation}
where $(p,p') = e,o$ and 
\begin{eqnarray}
C(\omega, \omega')&=& { 1 \over \omega - \epsilon_f} - {1 \over \omega - \epsilon_f - U_f}  \nonumber \\
&& +{ 1 \over \omega' - \epsilon_f}  - {1 \over \omega' - \epsilon_f - U_f} \;.
\end{eqnarray} 
In comparison to the notations in Ref.~\cite{Zhu06}, $J_e = J_{ee}$, $J_o=J_{oo}$ and $J_m = (J_{eo}+J_{oe})/2$. With $\epsilon=-U/2$ and $|\omega|$,$|\omega'| \ll U$, $C(\omega, \omega') \approx 8/U$. The Kondo coupling is taken as the value at the Fermi energy $J_K = (J_e^2+J_0^2+2J_m^2)^{1/2}/\sqrt{2} = 8\Gamma_0 /(\pi \rho_0 U)$. However, the energy dependence of the Kondo coupling is crucial to determine the generated RKKY interaction, which is 
\begin{equation}
I =  {\rho_0\over 2} \int_0^1 d\omega \int_{-1}^0 d\omega' { J_e^2 (\omega, \omega') + J_o^2(\omega,\omega') - 2J_m^2(\omega,\omega') \over \omega'-\omega}\;.
\end{equation}
If we take the hybridization functions to be constants,  we find that $I= -2\ln2 \rho_0 (J_e-J_o)^2$. It is then always ferromagnetic except when $J_e=J_o$, i.e., the two impurities located far away. When $\Gamma_{e,o}(\omega) = \Gamma_0 (1\mp \omega)$, we evaluate $I \approx 0.20 \rho_0 J_K^2$, i.e., antiferromagnetic as expected.


\begin{thebibliography}{99}
\bibitem{footnote1} We follow Ref. 1 for the definition of the coherence scale. But please notice that in many other literatures, the coherence scale is referred to a different energy scale, rather associated with the Fermi liquid behaviors of quasiparticles, or the Fermi liquid temperature $T_{L}$ in this paper.
\bibitem{footnote2} For ferromagnetic RKKY interactions, there are also two energy scales present, characterizing a two-stage Kondo process~\cite{Silva96}.
\end{thebibliography}
\end{document}